\documentclass[prl,twocolumn]{revtex4}
\usepackage{graphicx}
\renewcommand{\v}[1]{{\bf #1}}
\newcommand{\sign}{{\rm sign}}

\newcommand{\w}{{\omega}}

\def\eqa{\begin{eqnarray}}
\def\eea{\end{eqnarray}}
\newcommand{\eq}{\begin{equation}}
\newcommand{\ee}{\end{equation}}

\newcommand{\Det}{{\rm Det}}
\renewcommand{\Im}{{\rm Im}}

\newcommand{\ra}{\rightarrow}

\begin{document}

\title{Impurity and interface bound states in $d_{x^2-y^2}+id_{xy}$ and $p_x+ip_y$ superconductors}
\author{Qiang-Hua Wang$^{a}$ and Z. D. Wang$^{b}$} \affiliation{${(a)}$ National Laboratory
of Solid State Microstructures,Institute for Solid State Physics,
Nanjing University, Nanjing 210093, China}
\affiliation{${(b)}$Department of Physics,University of Hong Kong,
Pukfulam Road, Hong Kong, China}


\begin{abstract}
Motivated by recent discoveries of novel superconductors such as
Na$_x$CoO$_2\cdot y$H$_2$O and Sr$_2$RuO$_4$, we analysize
features of quasi-particle scattering due to impurities and
interfaces for possible gapful $d_{x^2-y^2}+id_{xy}$ and
$p_x+ip_y$ Cooper pairing. A bound state appears near a local
impurity, and a band of bound states form near an interface. We
obtained analytically the bound state energy, and calculated the
space and energy dependent local density of states resolvable by
high-resolution scanning tunnelling microscopy. For comparison we
also sketch results of impurity and surface states if the pairing
is nodal p- or d-wave.
\end{abstract}

\maketitle

Recently, Takada {\it et al} discovered a novel superconductor
Na$_x$CoO$_2\cdot y$H$_2$O ($x=0.35$) with a superconducting
transition temperature $T_c=5K$. \cite{takada} A few features of
this material bare strong connection to cuprates: 1) It has a
layered structure. 2) As $Cu^{2+}$ in cuprates, Co$^{4+}$ atom is
in a spin-1/2 state according to first principle calculation of
Singh. \cite{singh}  Combined with the fact that the cobalt
triangular lattice is frustrating to anti-ferromagnetic ordering,
the new material offers a likely situation for the physics of
Anderson's resonating valence bond (RVB) theory. \cite{anderson}
Soon after the discovery, theories based on RVB physics
\cite{baskaran,kumar,wll,ogata} and renormalization group analysis
\cite{honerkamp} predicted $d+id'$-wave pairing ($d=d_{x^2-y^2}$
and $d'=d_{xy}$), while other theories suggest $p_x+ip_y$-wave
pairing derived from the weak ferromagnetic instability,\cite{hu}
in close analogy to the case of Sr$_2$RuO$_4$.\cite{sro}
Identifying the pairing symmetry would be a necessary step toward
the understanding of the new superconductor. In this paper, we
propose tunnelling measurements of impurity and interface states
that are sensitive to both the gap amplitude and the internal
phase of the gap function. Such measurements have played
invaluable roles in high temperature superconductors in the
context of nodal d-wave pairing .\cite{tsuei,pandavis} Our main
results are as follows. As a consequence of the full gap as well
as the internal phase degrees of freedom of $d+id'$ and $p_x+ip_y$
Cooper pairs, a bound state appears at any nonzero scattering
intensity near a local impurity and a band of bound states form
near an interface. The bound state energy is near the gap edge at
weak scattering strength, and it approaches zero energy (the Fermi
level) at increasing scattering strengths. We also calculated the
energy and space dependent LDOS, whose rich features are directly
resolvable by future STM and can help identify the pairing
symmetry in the new superconductors. For comparison, we also
mention briefly the results of impurity and interface states for
nodal p- and d-wave pairing.

As usual the elastic scattering problem is best described in terms
of the retarded T-matrix formulation, \eq
G(i,j)=G_0(i,j)+\sum_{a,b}G_0(i,a)T(a,b)G_0(b,j),\label{G} \ee
where $a,b$ denotes the position of the impurities and all other
notations are standard. We suppressed the energy dependence in the
Green's functions, as it is conserved in elastic scattering. The
T-matrix is given by \eq
(T^{-1})(a,b)=(V^{-1})(a,b)-G_0(a,b),\label{T1} \ee where $V(a,b)$
is the general impurity potential that may be off-diagonal. In our
case, $G$, $G_0$ and $T$ are further $2\times 2$ matrices in the
particle-hole Nambu space. The scattering problem is solved once
$G_0$ is known. The LDOS at site $i$ is given by \eq
N(i,\w)=-\Im[G_{11}(i,i;\w)+G_{22}(i,i;-\w)]/\pi, \label{Ni} \ee
with the energy argument $\w$ restored. A peak in $N(i,\w)$
appears if either $G_{11}(i,i;\w)$ or $G_{22}(i,i;-\w)$ diverges.
This bound/resonance state occurs if $\Det[T^{-1}(\pm \w)]=0$. Due
to the mixing of particle and hole in the presence of pairing, it
is possible that there are two peaks in $N(i,\w)$ but
$\Det(T^{-1})=0$ is satisfied at only one energy, or vice versa.
In the following discussion, we always count the bound/resonance
states according to the peaks seen in the total density of states
$N(i,\w)$.

Let us write the gap function as , in the momentum space,
$\Delta_\v k=\Delta e^{il\theta_\v k}$ where $\Delta$ is the gap
amplitude, $\theta_\v k $ is the azimuthal angle of the vector $\v
k$ and $l=0, \pm 1,\pm 2$ for gapful s-, p- and d-wave pairing,
respectively. We include the case of s-wave pairing for
comparison. The above pairing function is of simplified form,
suitable near the normal state Fermi surface, and suffices for
qualitative discussion of low energy quasi-particle states. Then
$G_0(i,j)=G_0(\v r)$ (with $\v r=\v r_i-\v r_j$) is given by \eqa
& & G_0(\v r)=\int \frac{d^2\v k}{(2\pi)^2}
\frac{\w_+\sigma_0+\epsilon_k\sigma_3+\Delta\sum_\nu e^{ i\nu
l\theta_\v k}\sigma_\nu}{\w_+^2-\epsilon_k^2-\Delta^2}e^{i\v
k\cdot \v r}\nonumber\\& &\sim -\frac{\pi N_0\Big(\w_+J_0(k_F
r)\sigma_0 +\Delta J_l(k_F r)\sum_\nu e^{i\nu l\theta_{\v
r}}\sigma_\nu\Big)}{\sqrt{\Delta^2-\w_+^2}}, \label{G0} \eea where
$\w_+=\w+i0^+$, $\nu=\pm$, $\epsilon_k$ is the normal state energy
dispersion, $\sigma_0$ is the $2\times 2$ unit matrix,
$\sigma_{1,2,3}$ are the Pauli matrices,
$\sigma_{\pm}=(\sigma_1\pm i\sigma_2)/2$ and $
J_l(u)=\int_0^{2\pi}d\theta\cos(l\theta)\exp(iu\cos\theta)/2\pi$
is the Bessel function. In arriving at the above results, we have
assumed a cylindrical Fermi surface with Fermi vectors of
magnitude $k_F$, and constant density of states $N_0$ near the
Fermi level. We emphasize that a particle-hole asymmetry is
present in the normal state DOS of Na$_x$CoO$_2$. We shall comment
on such effects without going into details in the following
qualitative discussions.

{\bf \em I. Scattering from a local impurity:} In this case we set
the impurity site at the origin, i.e., $a=b=0$, and drop these
indices in $V=V_m\sigma_0+V_s\sigma_3$ and
$T^{-1}=V^{-1}-G_0(0,0)$, where $V_{m,s}$ is the strength of
(classical) magnetic/scalar potential. With $G_0$ in Eq.(\ref{G0})
at hand, the T-matrix is now given by \eq T^{-1}=V^{-1}+\frac{\pi
N_0}{\sqrt{\Delta^2-(\w_+)^2}}\Big(\w_+\sigma_0+\Delta
\delta_{0l}\sigma_1\Big).\label{Timp}\ee One sees that $\Im
(T^{-1})\ra 0$ in the sub-gap regime $\w^2<\Delta^2$, so that a
true {\em bound} state could be generated in this regime provided
$\Det (T^{-1})=0$. This should be contrasted to the case of
virtually bound impurity state, or the resonant impurity state, in
the case of nodal d-wave pairing. \cite{balatsky} The condition
$\Det (T^{-1})=0$ is governed by the value of the dimensionless
scattering strengths $c_{m,s}=\pi N_0 V_{m,s}$. A few cases are
classified as follows.

The case of s-wave pairing ($l=0$) has been discussed in the
literature,\cite{yu,morr} and we list briefly some of the results
for comparison: 1) No sub-gap bound states can be generated for a
scalar impurity ($V_s\neq 0$ and $V_m=0$). This is consistent with
the Anderson theorem that s-wave pairing is robust against scalar
impurities. 2) A magnetic impurity ($V_s=0$ and $V_m\neq 0$) can
always generate a pair of bound states at energies $\w_b=\pm
|\Delta|(1-c_m^2)/(1+c_m^2)$. In any case, the spatial modulation
of the LDOS in the presence of the impurity is given by $
N(i,\w)-{\cal N}(\w)\propto J_0^2(k_F r_i)$ from Eqs.(\ref{G}),
(\ref{G0}) and (\ref{Timp}) under the given approximation. Here
${\cal N}(\w)$ is the site-independent bulk density of states in
the superconducting state.

\begin{figure}
\includegraphics[width=8.5cm]{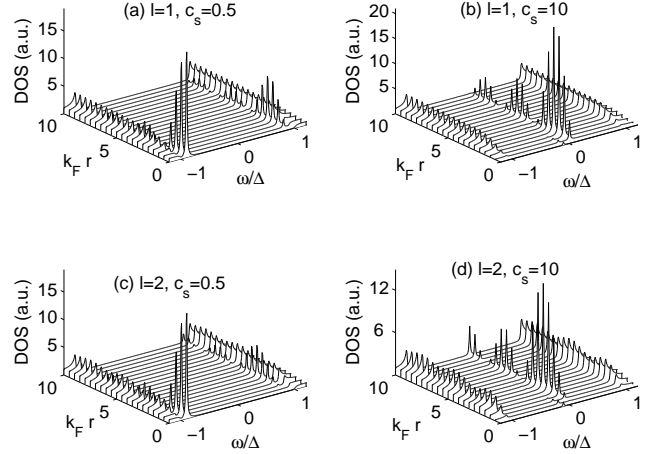}
\caption{Density of states as a function of energy $\w$ and the
radial distance $r$ off a scalar impurity. See the text for
details.}
\end{figure}
\begin{figure}
\includegraphics[width=8.5cm]{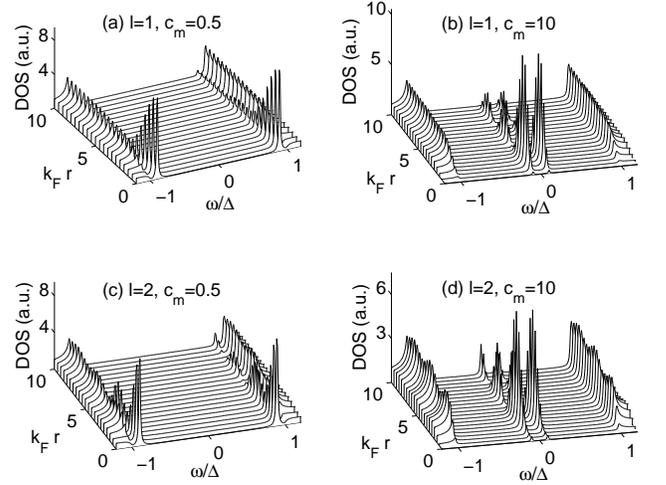}
\caption{The same plot as Figs.1 but for a magnetic impurity.}
\end{figure}

For p- and d-wave pairing, the off-diagonal $\sigma_1$-component
in $T^{-1}$ is zero. This is not an accidental result from the
adopted approximation, but rather a rigorous result from the
pairing symmetry, which forbids the on-site pairing amplitude
(related to the anomalous part of $G_0(0)$) to be finite.
Consequently, both scalar and magnetic impurities can generate
bound states. 1) For a scalar impurity, $\Det(T^{-1})=0$ is
satisfied at \eq \w_b=\pm
|\Delta|/\sqrt{1+c_s^2}.\label{wbGapScalarImp}\ee In general this
implies two peaks in the LDOS according to Eq.(\ref{Ni}). However,
depending on the ratio $\w_b^2/\Delta^2$ one of the peaks may
dominate over the other, with an associated change in the spatial
dependence of the LDOS. We present a few examples of the energy
and space dependent LDOS in Figs.1 for Cooper pairing with
$l=1,2$. For the weak impurity case $c_s=0.5$ in Figs.1 (a) and
(c), $\w_b$ is near the gap edge, the dominant peak is at the
energy with opposite sign to $c_s$, and the corresponding DOS
right at the impurity site is maximal. In contrast, for the strong
impurity case $c_s=10$ in Figs.1 (b) and (d), $\w_b$ is
approaching the Fermi level (zero energy), the dominant DOS peak
energy has the same sign as that of $c_s$, and the corresponding
DOS is vanishing right at the impurity site. Note that the
cusp-like feature at $\w=\pm \Delta$ away from the impurity is
just a feature of ${\cal N}(\w)$. 2) For a magnetic impurity,
$\Det(T^{-1})=0$ is satisfied at \eq
\w_b=-\sign(c_m)|\Delta|/\sqrt{1+c_m^2}\label{wbGapMagneticImp}.
\ee Since $T^{-1}\propto \sigma_0$ in this case, there are
actually two peaks in DOS, according to Eq.(\ref{Ni}), located
symmetrically with respect to the Fermi level. Examples are shown
in Figs.2, in comparison to Figs.1. By inspection, we see that
except for the symmetrical peaks, Figs.2 are basically similar to
Figs.1. On the other hand, in both scalar and magnetic impurity
cases the difference between $l=1$ and $l=2$ is mild. This would
pose difficulty for STM to resolve this quantum number.
Fortunately this can be resolved easily by other means such as
spin susceptibility measurements from the fact that singlet
pairing ($l=2$ here) forms a gap for spin excitations while
triplet pairing ($l=1$ here) does not.

It is pertinent at this stage to comment on the effect of
particle-hole asymmetry in the normal state Fermi surface. As can
be seen from the derivation of $G_0$, this would introduce a
$\sigma_3$-component in $G_0(0,0)$, which effectively acts as an
excess energy-dependent scalar potential in $T^{-1}$. Therefore,
the effect is to modify the bound state energy, and to break the
symmetry of the bound state energies in the case of magnetic
scattering.

{\bf \em II. Scattering from an interface:} We shall model an
interface by an extended line of impurities. This could be
fabricated by chemical erosion. It is to our advantage in that the
T-matrix formalism can still be applied. Since the unperturbed
system at hand has rotational symmetry, the interface states
should not depend on the surface normal direction $\hat{n}$, which
we fix to be $\hat{n}=\hat{y}$ for definiteness. Due to the
remaining translation symmetry along the $x$-axis, we can do
partial Fourier transforms of Eqs.(\ref{G}), (\ref{T1}) and
(\ref{G0}) with respect to $x$ to find the reduced t-matrix
equations at the $x$-direction wave vector $q$ as
\eqa & &g(y_i,y_j)=g_0(y_i-y_j)+g_0(y_i)tg_0(-y_j),\nonumber\\
& &g_0(y)\sim \frac{2\pi N_0\Big(\w_+\cos
py\sigma_0+\Delta\sum_\nu\cos(py+\nu l\theta_q)\sigma_\nu\Big)}
{-p\sqrt{\Delta^2-\w_+^2}}, \nonumber\\
& & t^{-1}=v^{-1}+\frac{2\pi
N_0}{p\sqrt{\Delta^2-\w_+^2}}(\w_+\sigma_0+\Delta\cos
l\theta_q\sigma_1),\nonumber \eea where
$v=V_m\sigma_0+V_s\sigma_3$ is the same as the form of a single
impurity, $p=\sqrt{k_F^2-q^2}=k_F|\sin\theta_q|$ and
$\theta_q=\arccos(q/k_F)$. The conserved momentum $q$ is
suppressed in the arguments of $g$, $g_0$ and $t^{-1}$ for
brevity. The problem is reduced to an effective single impurity
scattering in one-dimension. With the implicit $\w$ and $q$
arguments restored, the partial DOS is given by $ N(\w,y;q)=-\Im
[g_{11}(\w,y;q)+g_{22}(-\w,y;q)]/\pi$, and the total density of
states is $N(\w,\v r)=\int dqN(\w,y;q)/2\pi$, which is independent
of $x$ due to the translation symmetry. (The integration over $q$
should be cutoff at $\pm k_F$.) Again $\Det (t^{-1})=0$ would
predict a bound state.

Although the above formulation is versatile to deal with any value
of $V_{m,s}$, we shall consider only the more likely scalar
interface with $V_m=0$. It is easy from the above equations that
bound states occur at energies given by \eq \w_b=\pm
\Delta\sqrt{\frac{4c_s^2\cos^2l\theta_q+k_F^2\sin^2\theta_q}
{4c_s^2+k_F^2\sin^2\theta_q}},\label{wbGapSurf}\ee which clearly
form two bands. (Note that we have taken the lattice constant to
be unity so that $k_F$ is dimensionless.) It is also clear that no
sub-gap bound states exist for s-wave pairing ($l=0$).

\begin{figure}
\includegraphics[width=8.5cm]{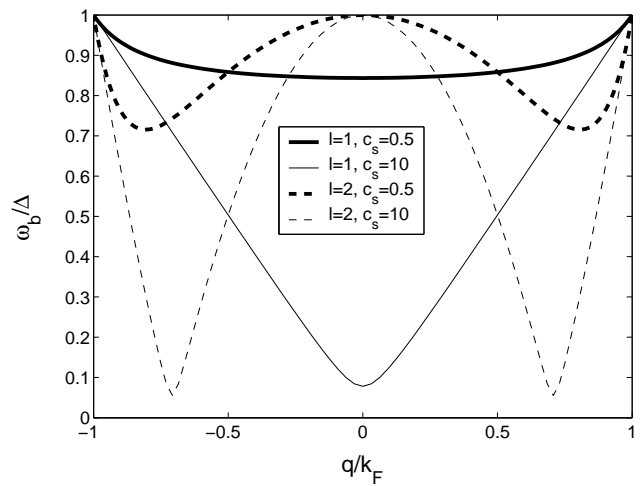}
\caption{Dispersion of the positive bound state energy as a
function of the wave vector along the interface. See the text for
details.}
\end{figure}

\begin{figure}
\includegraphics[width=8.5cm]{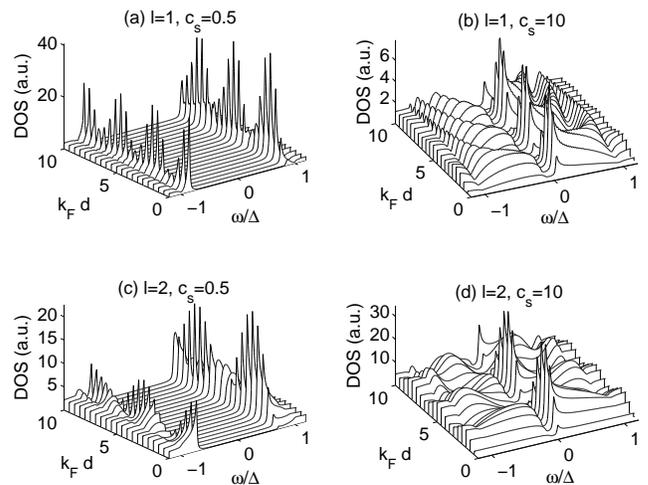}
\caption{Density of states as a function of energy $\w$ and the
distance $d$ off an interface. See the text for details.}
\end{figure}

The dispersion of the positive bound state energy is plot in Fig.3
for weak ($c_s=0.5$, thick lines) and strong ($c_s=10$, thin
lines) interface with p-wave ($l=1$, solid lines) and d-wave
($l=2$, dashed lines) pairings. Here we have set $k_F=\pi/2$ for
calculation. One sees that the energy disperses near the gap edge
for weak scattering interfaces, and it tends to cover the whole
sub-gap regime for strong interface scattering. Furthermore, the
difference between p- and d-wave pairing is reflected in the
number of minima, being identical to $l$, in the dispersion.

The spatial dependence of the LDOS near the interface can be
calculated from the above theory. Examples are shown in Figs.4.
Consistent with the above bound state energy dispersion, the
sub-gap states are near the gap edge (or tend to cover the whole
gap regime) for a weak (or strong) scattering interface,
distributed more or less symmetrically (or asymmetrically) with
respect to the Fermi level. In the limit of unitary scattering
$c_s\ra \infty$ (not shown here) the LDOS becomes symmetrical in
energy again. An interesting feature in Figs.4(b) and (d) is that
the peaks or bumps in energy oscillate with increasing distance
from the interface, forming wave-like pattern in the
energy-distance space. It is also clear from Figs.4 that the
spatial profile decays much more slowly than the single impurity
cases in Figs.1 and 2.

We note that Matsumoto and Sigrist \cite{sigrist} have addressed
the quasi-particle states near a sample surface and a topological
domain wall (with a $\pi$ phase shift in the pairing gap) for
$p_x+ip_y$-wave pairing in terms of quasi-classical theory. They
found that sub-gap states appear near the domain wall but not the
surface. Our interface is actually a non-topological domain wall
but with potential scattering.

{\em III. The case of nodal p- and d-wave pairing:} Along similar
lines to that sketched above, we have also considered the impurity
and interface states for nodal p- and d-wave pairing for
comparison. The gap function may be written as $\Delta_\v
k=\Delta\sin l\theta_\v k$ or $\Delta_\v k=\Delta\cos l\theta_\v
k$, depending on whether one of the nodal or antinodal directions
is along the $x$-axis. Note that there are only one nodal and one
antinodal direction for p-wave pairing. Due to limited space we
sketch the results without going into details.

For the local impurity case, we found resonant energies at
$\w_r\sim\pm \pi\Delta/[lc_{m,s}\ln (4lc_{m,s}/\pi)]$ for a
scalar/magnetic impurity. This reduces to the known result in the
case of nodal d-wave pairing ($l=2$).\cite{balatsky} The new
features for the case of nodal p-wave pairing is that that LDOS
pattern near the impurity is {\em two-fold} symmetric, forming
stripe-like features extending along the anti-nodal direction,
which should be compared to the four-fold symmetric pattern in the
case of nodal d-wave pairing.\cite{balatsky,wl,wang,zhang}

On the other hand, the interface states depend on the interface
orientation: 1) If the (scalar scattering) interface is along one
of the nodal directions, say $\hat{x}$, there are bound states at
energies $\w_b=\pm \Delta|k_F\sin \theta_q\sin l
\theta_q|/\sqrt{4c_s^2+k_F^2\sin^2\theta_q}$. The definition of
$\theta_q$ is the same as in section II. These energies all
approach zero in the unitary limit $c_s\ra \infty$. The abundance
of zero energy states is due to the fact that in this limit
quasi-particles reflect spectacularly from the interface,
experiencing a sign change of the gap. The same physics is nicely
described in Refs.\cite{hucr,sheehy} in other contexts. 2) Finally
if the interface is along one of the anti-nodal directions,
redefined also as $\hat{x}$, there are resonant states exactly at
$\w_r=\pm \Delta\cos l\theta_q$ irrespectively of the scattering
strength. In fact this is equivalent to the case of s-wave pairing
but with a $q$-dependent gap amplitude.

{\em IV. Closing remarks:} We have only shown results for the
cases $l=\pm 1,\pm 2$ that are relevant in the new
superconductors, but the theory is clearly general for any integer
value of $l$. There are some details missing in the theory,
however. First, it does not take into account possible anisotropy
in the normal state Fermi surface. For example, in Na$_x$CoO$_2$,
the Fermi surface has a rounded hexagonal structure.\cite{singh}
Such anisotropy may cause corresponding anisotropic LDOS pattern
around impurities. Second, Eq.(\ref{G0}) is obtained by fixing the
momentum on the Fermi surface while integrating over energy. This
possibly leaves out an excess decay of $G_0$ in space with the
length scale $\xi=v_F/\Delta$. Apart from such details, our
qualitative analytical results are robust.

\acknowledgments{This was supported by NSFC 10204011 and 10021001,
the Ministry of Science and Technology of China
(NKBRSF-G1999064602), and the RGC grant of Hong Kong.}

\end{document}